\documentclass[aps,prl,twocolumn,superscriptaddress,nofootinbib]{revtex4-1}

\usepackage[colorlinks=true,citecolor=red,urlcolor=blue]{hyperref}
\usepackage{amssymb}
\usepackage{amsmath,color}
\usepackage{graphicx}
\usepackage{bbm}
\usepackage{verbatim}
\usepackage[T1]{fontenc}
\usepackage[utf8]{inputenc}
\usepackage[normalem]{ulem}

\usepackage{url}
\usepackage{xcolor}
\usepackage{caption}
\usepackage{subcaption}

\usepackage{diagbox}
\usepackage{color}
\usepackage{multirow}
\usepackage{hhline}
\usepackage{tabularx}

\begin{document}


\title{Eccentricity of Long Inspiraling Compact Binaries Sheds Light on Dark Sirens}


\author{Tao Yang}
\email[]{yangtao.lighink@gmail.com}
\affiliation{Center for the Gravitational-Wave Universe, Astronomy Program Department of Physics and Astronomy, Seoul National University, 1 Gwanak-ro, Gwanak-gu, Seoul 08826, Korea}

\author{Rong-Gen Cai}
\email[]{cairg@itp.ac.cn}
\affiliation{CAS Key Laboratory of Theoretical Physics, Institute of Theoretical Physics, Chinese Academy of Sciences, Beijing 100190, China}
\affiliation{School of Physical Sciences, University of Chinese Academy of Sciences, No. 19A Yuquan Road, Beijing 100049, China}
\affiliation{School of Fundamental Physics and Mathematical Sciences, Hangzhou Institute for Advanced Study (HIAS), University of Chinese Academy of Sciences, Hangzhou 310024, China}

\author{Zhoujian Cao}
\email[]{zjcao@bnu.edu.cn}
\affiliation{Department of Astronomy, Beijing Normal University, Beijing 100875, China}
\affiliation{School of Fundamental Physics and Mathematical Sciences, Hangzhou Institute for Advanced Study (HIAS), University of Chinese Academy of Sciences, Hangzhou 310024, China}

\author{Hyung Mok Lee}
\email[]{hmlee@snu.ac.kr}
\affiliation{Astronomy Program Department of Physics and Astronomy, Seoul National University, 1 Gwanak-ro, Gwanak-gu, Seoul 08826, Korea}

\date{\today}

\begin{abstract}
The localization and distance inference of gravitational waves are two crucial factors for dark sirens as precise probes of cosmology, astrophysics, and fundamental physics. In this Letter, for the first time we investigate the parameter estimation of gravitational waves emitted by the eccentric compact binaries in the mid-frequency (0.1--10 Hz) band. Based on the configuration of one cluster of DECIGO (B-DECIGO), we simulate five types of typical compact binaries in GWTC-3 with component mass ranging from $\mathcal{O}(1\sim100)~M_{\odot}$. For each type of binaries, we assign discrete eccentricities from 0 to 0.4 at 0.1 Hz in $10^3$ random orientations. The multiple harmonics induced by eccentricity can break the degeneracy between parameters. We find that with eccentricity $e_0=0.4$, these typical binaries can achieve $\mathcal{O}(10^2-10^4)$ improvement for the distance inference in the near face-on orientations, compared to the circular case. More importantly, a nonvanishing eccentricity ($0.01\sim0.4$) can significantly improve the source localization of the typical binary black holes, most by $1.5\sim{3.5}$ orders of magnitude. Our result shows the remarkable significance of eccentricity for dark sirens in the mid-band as precise probes of the Universe.
\end{abstract}

\maketitle

\textit{Introduction.}--The discovery of gravitational waves (GWs) provides us with a novel probe to the Universe~\cite{LIGOScientific:2021djp}.  
In particular, the discovery of binary neutron star (BNS) GW170817~\cite{LIGOScientific:2017vwq} and the kilonova emission from its remnant~\cite{Coulter:2017wya,LIGOScientific:2017ync,LIGOScientific:2017zic} provided the first GW standard siren measurement of the Hubble constant~\cite{LIGOScientific:2017adf}.  However, for GWs without confirmed electromagnetic (EM) counterparts, the host galaxies and their redshifts cannot be determined directly. We call this type of GWs the ``dark sirens'' as opposed to ``bright sirens'' which have EM counterparts. Up to now, only one bright siren (GW170817) has been confirmed. Dark sirens detected so far have also been used to measure the Hubble constant by determining the source redshifts in a statistical way~\cite{DES:2019ccw,LIGOScientific:2019zcs,DES:2020nay,LIGOScientific:2021aug}. Until now, the accuracy gain in Hubble constant with dark sirens is very limited due to the uncertainty of the host galaxies (and hence its redshift) within the localized region.

GW standard siren, though currently not precise enough, is one of the most promising probes to arbitrate the current tension between local and high-$z$ measurements of the Hubble constant~\cite{Freedman:2017yms,Verde:2019ivm,Borhanian:2020vyr,Chen:2017rfc}, as well as study the feature of dark energy and test general relativity~\cite{Dalal:2006qt,Zhao:2010sz,Cai:2016sby,Belgacem:2019tbw,Finke:2021aom,Yang:2021qge}. To fully utilize the potential of standard sirens on cosmology, astrophysics, and fundamental physics, the unbiased and precise inferences of luminosity distance and redshift of the source are very essential. 
The determination of the redshift has recourse to either the EM counterpart (for bright sirens) or statistical method by counting all the potential host galaxies within the localized region (for dark sirens). The latter heavily depends on the precise source localization. Considering the fact that bright sirens are much rarer than dark sirens in current detections and even in the forecasted catalogs based on future detector networks~\cite{Belgacem:2019tbw,Yang:2021qge}, the techniques to improve the estimation of distance and sky localization parameters for dark sirens are very crucial. In this Letter, we demonstrate that the eccentricity of the compact binaries plays an important role in the distance inference and localization for dark sirens.

Many investigations suggest compact binaries that emit GWs can have non-negligible eccentricities and may contribute observational features in the sensitivity band of ground and space-based detectors~\cite{Antonini:2012ad,Samsing:2013kua,Thompson:2010dp,East:2012xq}. 
Some studies indicate that a fraction of the binaries possess eccentricities larger than 0.1 at 10 Hz~\cite{Wen:2002km,Silsbee:2016djf,Antonini:2017ash,Liu:2019gdc}. Recent works~\cite{Sun:2015bva,Ma:2017bux,Pan:2019anf} showed that, for the case of $100~M_\odot$ binary black hole (BBH), eccentricity can improve the source localization by a factor of 2 in general with the ground-based detector networks. However, such an improvement is not sufficient to considerably reduce the uncertainty of the host galaxy identification for LIGO-Virgo's BBH since there are typically thousands of potential host galaxies within their sky locations. 

In this Letter, we make the investigation for eccentric binaries in the mid-frequency band, i.e., 0.1-10 Hz, where the space-based detector like DECIGO is most sensitive~\cite{Kawamura:2020pcg}. The reasons for carrying out such an analysis in the mid-band are as follows. Most analyses for current GW detections do not find strong evidence of the eccentricity in the LIGO-Virgo band~\cite{Lower:2018seu,LIGOScientific:2019dag,Romero-Shaw:2019itr,Wu:2020zwr,LIGOScientific:2021tfm}, except for a very recent report of highly eccentric merger GW190521 with $e=0.69^{+0.17}_{-0.22}$~\cite{Gayathri:NA} (see also~\cite{Romero-Shaw:2020thy}). At lower frequencies, we have much higher possibility to detect the orbital eccentricity of the binaries. In addition, the compact binaries observed by the mid-band detectors have long inspiral phase (days to years). For the mid-frequency, the detectors need to be deployed in space in order to avoid seismic and Newtonian noise in the ground. The space-borne detector changes position significantly during the long-time observation, causing the antenna response function to change with time. The movement of the detector can provide a much more precise angular resolution of GW sources. The long-time observation increases the signal-to-noise ratio (SNR) which can further improve the parameter estimation. These factors pose the following question that has not yet been addressed in the literature: can eccentricity improve the measurement of distance and localization of the compact binaries observed by a space-borne detector in the mid-frequency band? In this Letter, we attempt to answer this question by mocking up some types of typical compact binaries. 

\textit{Methodology.}--To construct the eccentric waveform, we adopt the non-spinning, inspiral-only EccentricFD waveform approximant available in {\sc LALSuite}~\cite{lalsuite}. We use {\sc PyCBC}~\cite{alex_nitz_2022_6324278} to generate the waveform. EccentricFD corresponds to the enhanced post-circular (EPC) model~\cite{Huerta:2014eca}. To the zeroth order in the eccentricity, the model recovers the TaylorF2 post-Newtonian (PN) waveform at 3.5 PN order~\cite{Buonanno:2009zt}. To the zeroth PN order, the model recovers the PC expansion of~\cite{Yunes:2009yz}, including eccentricity corrections up to order $\mathcal{O}(e^8)$. In the eccentric waveform, we have 11 parameters, $P=\{\mathcal{M}_c,\eta,d_L,\iota,\theta,\phi,\psi,t_c,\phi_c,e_0,\beta\}$, with two additional parameters \{$e_0,~\beta$\} to the circular TaylorF2 model. Here, $e_0$ is the initial eccentricity at frequency $f_0$ and $\beta$ is the azimuthal component of inclination angles (longitude of ascending nodes axis). 
The EPC model can be written as~\cite{Huerta:2014eca}
\begin{equation}
\tilde{h}(f)=-\sqrt{\frac{5}{384}}\frac{\mathcal{M}_c^{5/6}}{\pi^{2/3}d_L}f^{-7/6}\sum_{\ell=1}^{10}\xi_{\ell}\left(\frac{\ell}{2}\right)^{2/3}e^{-i\Psi_{\ell}} \,.
\end{equation} 
When $e_0=0$ it recovers the circular TaylorF2 model ($\ell=2$). $\xi_{\ell}$ are functions of $e_0$ and angular parameters $P_{\rm ang}=\{\iota,~\theta,~\phi,~\psi,~\beta\}$~\cite{Yunes:2009yz}. The eccentricity adds more harmonics to the waveform. Multiple harmonics make the distance and angular parameters nontrivially coupled, enabling us to break the degeneracy among these parameters.

In this Letter, we consider only one cluster of DECIGO. This configuration is similar to B-DECIGO which is a scientific pathfinder of DECIGO and will be launched much earlier. Our results can serve as a forecast for one cluster of DECIGO as well as for the early lunched B-DECIGO. We follow~\citet{Rubbo:2003ap} for the modeling of the space-borne detectors, with the arm length $L=1000$ km~\cite{Kawamura:2020pcg}.  For the sensitivity curve,
we use the fitting formula in~\cite{Yagi:2011wg} but rescale it according to the latest white paper of DECIGO~\cite{Kawamura:2020pcg}.

To estimate the uncertainty and covariance of the waveform parameters, we adopt the Fisher matrix technique 
$\Gamma_{ij}=\left(\frac{\partial h}{\partial P_i},\frac{\partial h}{\partial P_j}\right)$, with $P_i$ one of the 11 waveform parameters.
The inner product is defined as $(a,b)=4\int_{f_{\rm min}}^{f_{\rm max}}\frac{\tilde{a}^*(f)\tilde{b}(f)+\tilde{b}^*(f)\tilde{a}(f)}{2 S_n(f)}df$. Then the covariance matrix of the parameters is $C_{ij}=(\Gamma^{-1})_{ij}$, from which the uncertainty of each parameter $\Delta P_i=\sqrt{C_{ii}}$. The error of the sky localization is $\Delta \Omega=2\pi |\sin(\theta)|\sqrt{C_{\theta\theta}C_{\phi\phi}-C_{\theta\phi}^2}$~\cite{Cutler:1997ta}. We calculate the partial derivatives $\partial \tilde{h}/\partial P_i$ numerically by $[\tilde{h}(f,P_i+dP_i)-\tilde{h}(f,P_i)]/dP_i$, with $dP_i=10^{-n}$.  For each parameter, we need to optimize $n$ to make the derivative converged so that the Fisher matrix calculation is reliable. To check the robustness of our methodology, we first adopt EccentricFD waveform with $e_0=0$ and check its consistency with the analytical TaylorF2 waveform. We find that after optimizing $n$ in the partial derivatives, the Fisher matrix calculation from the waveform given by {\sc PyCBC} and the analytical TaylorF2 are very consistent with each other. This consistency check paves the way for adopting the EccentricFD waveform in the nonvanishing $e_0$ cases.  

To quantify the influence of eccentricity on dark sirens in the mid-frequency band, we carry out this research by a specific strategy. We mock up five types of typical events from GWTC-3, i.e., a GW170817-like BNS with $(m_1,m_2)=(1.46,1.27)~M_{\odot}$, a GW200105-like neutron star–black hole binary (NSBH) with $(9.0,1.91)~M_{\odot}$, a GW191129-like light-mass BBH with $(10.7,6.7)~M_{\odot}$, a GW150914-like medium-mass BBH with $(35.6,30.6)~M_{\odot}$, and a GW190426-like heavy-mass BBH with $(106.9,76.6)~M_{\odot}$. Note that the light, medium, and heavy mass are in terms of the stellar-mass binaries in GWTC-3. The redshifts (distances) are also consistent with the real events.  To count the influence of the source orientation, we sample 1000 random sets of the angular parameters $P_{\rm ang}$ from the uniform and isotropic distribution. Since EPC waveform is tested valid with initial frequency up to 0.4~\cite{Huerta:2014eca}, for each event in every orientation, we assign six discrete eccentricities from 0 to 0.4 at $f_0=0.1$ Hz, i.e, $e_0=0$, 0.01, 0.05, 0.1, 0.2, and 0.4.  We would like to assess the effect of eccentricity on parameter estimation in different source orientations. Without loss of generality, we fix the coalescence time and phase to be $t_c=\phi_c=0$. We then totally have $5\times 6\times 1000=3\times10^4$ cases and for each case we calculate its SNR and Fisher matrix. 
We set the initial observational time to be around 1 year. Thus, we truncate the contribution of all $\ell$ harmonics that appear earlier than 1 year before the upper frequency (10 Hz). The strain is multiplied by the unit step function, $\tilde{h}_{\rm 1~yr}(f)=\tilde{h}(f)\mathcal{H}(2f-\ell f_{\rm start})$, with $\mathcal{H}(x)=1$ for $x\geq0$ and 0 otherwise~\cite{Yunes:2009yz}. The starting frequency of quadrupole mode $f_{\rm start}$ is set to be 0.2, 0.1, 0.059, 0.026, and 0.0105 Hz for the typical BNS, NSBH, light BBH, medium BBH, and heavy BBH, respectively. We use the frequency-rescaled detector response functions for different $\ell$ harmonics. The higher harmonics enter the detector band earlier. To derive the time-dependent detector response functions, we numerically solve the orbital phase evolution in~\cite{Yunes:2009yz} to obtain the time to coalescence $t(f)$ for the nonvanishing $e_0$. The time to coalescence at a specific frequency is smaller for a larger eccentricity.

We collect the results of the Fisher matrix for all of the $3\times10^4$ cases. For each typical event with a specific orientation, we define the ratios $R_{\Delta d_L}=\frac{\Delta d_L|_{e_0={\rm nonzero}}}{\Delta d_L|_{e_0=0}}$ and $R_{\Delta \Omega}=\frac{\Delta \Omega |_{e_0={\rm nonzero}}}{\Delta\Omega|_{e_0=0}}$, to show the improvement induced by eccentricity in that orientation. If $R<1$, there is an improvement in the relevant parameter. A smaller $R$ indicates a tighter constraint, and hence a larger improvement. We show the scatter plots of  $\Delta d_L/d_L$, $R_{\Delta d_L}$, $\Delta \Omega$, and $R_{\Delta \Omega}$ against $P_{\rm ang}$. We use $\iota$ to represent $P_{\rm ang}$ since the results are more sensitive to $\iota$ than other parameters. To give the statistical results, we define the minimum, mean, and maximum value of $x$ in the 1000 orientations as $\min(x)$, $\mathbb{E}(x)$, and $\max(x)$, respectively.

\begin{figure*}
\includegraphics[width=0.45\textwidth]{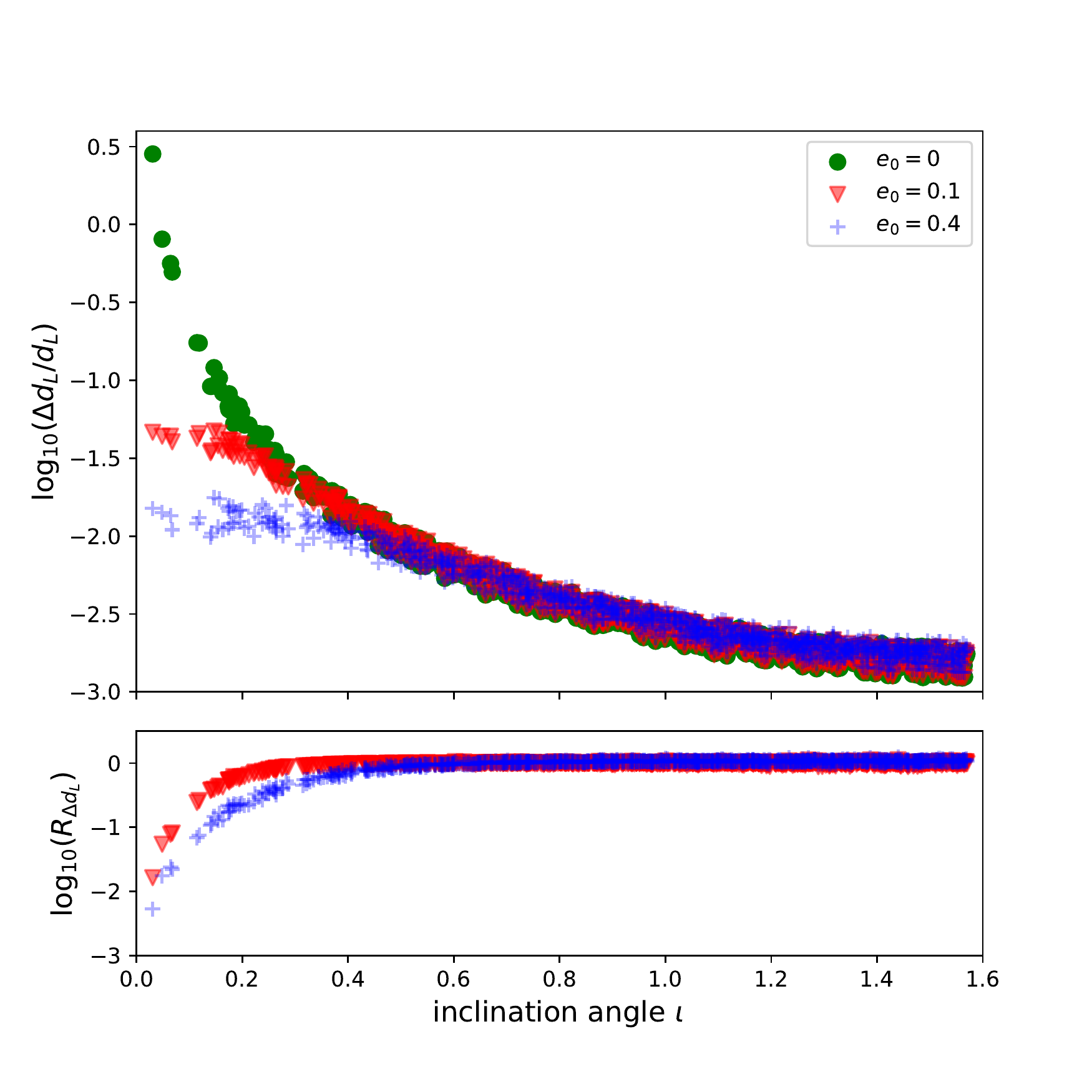} 
\includegraphics[width=0.45\textwidth]{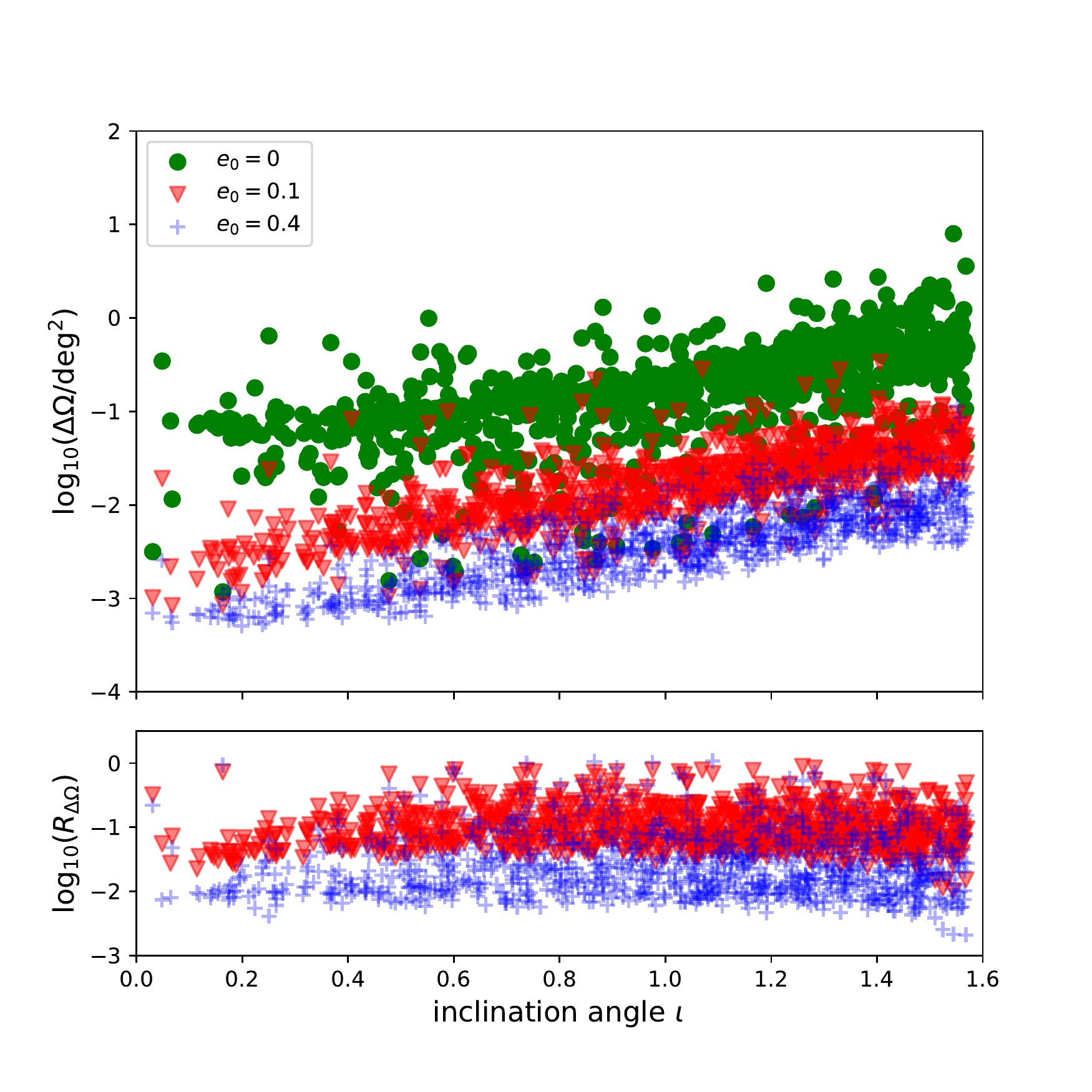} 
\caption{\textbf{Left:} The inference of luminosity distance against inclination angle for the GW170817-like BNS case. \textbf{Right:} The source localization against inclination angle for the GW150914-like medium-mass BBH case. }
\label{fig:result}
\end{figure*}

\textit{Distance inference and source localization.}--We select the distance inference of GW170817-like BNS and localization of GW150914-like medium BBH to represent the main features in our results, as shown in Fig.~\ref{fig:result}. 
In the near face-on orientations ($\iota\sim 0$) with $e_0=0$, luminosity distance largely degenerates with the inclination angle, hence $\Delta d_L/d_L$ is uncontrollably large. A nonvanishing eccentricity can break this degeneracy and significantly reduce the errors.  As shown in the left panel of Fig.~\ref{fig:result}, the $\max(\Delta d_L/d_L)$ of GW170817-like BNS is reduced from 2.84 ($e_0=0$) to $4.72\times10^{-2}$ ($e_0=0.1$) and $1.77\times10^{-2}$ ($e_0=0.4$). The largest improvement corresponds to around 61 and 189 times stricter. The huge improvement in the near face-on orientations holds true for all the typical events. We also find that the binary with a larger component mass and greater eccentricity can achieve more improvement. For GW190642-like heavy-mass BBH with $e_0=0.4$, it can most achieve $6\times10^3$ times improvement. For the typical BNS, NSBH and light BBH cases, the distance inference of circular and eccentric cases converges ($\mathbb{E}(R_{\Delta d_L})\sim1$) in the large inclination angle, where the degeneracy between $d_L$ and $\iota$ is relatively smaller and hence of the error. However, due to the significant improvement in the near face-on orientations, the $\mathbb{E}(\Delta d_L/d_L)$, which is of order $\mathcal{O}(10^{-2})$, is reduced by a factor of 2--4 with $e_0=0.1$. For the medium and heavy BBH cases, we find an overall improvement of distance inference even for the large $\iota$. The $\mathbb{E}(\Delta d_L/d_L)$ of heavy BBH is from 0.842 ($e_0=0$) to 0.123 ($e_0=0.1$) and 0.0565 ($e_0=0.4$), corresponding to 7--15 times improvement. For the improvement in a single orientation, we find $\mathbb{E}(R_{\Delta d_L})\sim 0.564~(e_0=0.1)$ and $0.290~(e_0=0.4)$. 

For the source localization of these typical events, we find eccentricity can lead to significant improvement only for the BBH cases which have larger component masses than BNS and NSBH cases.  As shown in the right panel of Fig.~\ref{fig:result}, the source localization of GW150914-like medium BBH is significantly improved in almost all orientations. Specifically, $\mathbb{E}(\Delta \Omega)$ and $\max(\Delta \Omega)$ are reduced from 0.290 and $7.97~\rm deg^2$ with $e_0=0$ to 0.0241 and $0.335~\rm deg^2$ with $e_0=0.1$, and to 0.00608 and $0.114~\rm deg^2$ with $e_0=0.4$. The average localization is improved by a factor of 12 ($e_0=0.1$) and 48 ($e_0=0.4$). In a single orientation with $e_0=0.1~(0.4)$, $\min(R_{\Delta \Omega})\sim1.02\times10^{-2}~(2.10\times10^{-3})$ corresponds to 100 (478) times for the largest improvement. Like the distance inference, a larger-mass compact binary with larger eccentricities can achieve more improvements for the source localization. For the heavy BBH case with $e_0=0.1~(0.4)$, $\mathbb{E}(\Delta \Omega)$ is shrunk from 157 to 32.6 (2.91) deg$^2$. The largest improvement is $\min(R_{\Delta \Omega})= 7.90\times 10^{-4}~(2.55\times 10^{-4})$, corresponding to $1.27\times10^3~(3.92\times10^3)$ times tighter.

\begin{figure}
\includegraphics[width=0.45\textwidth]{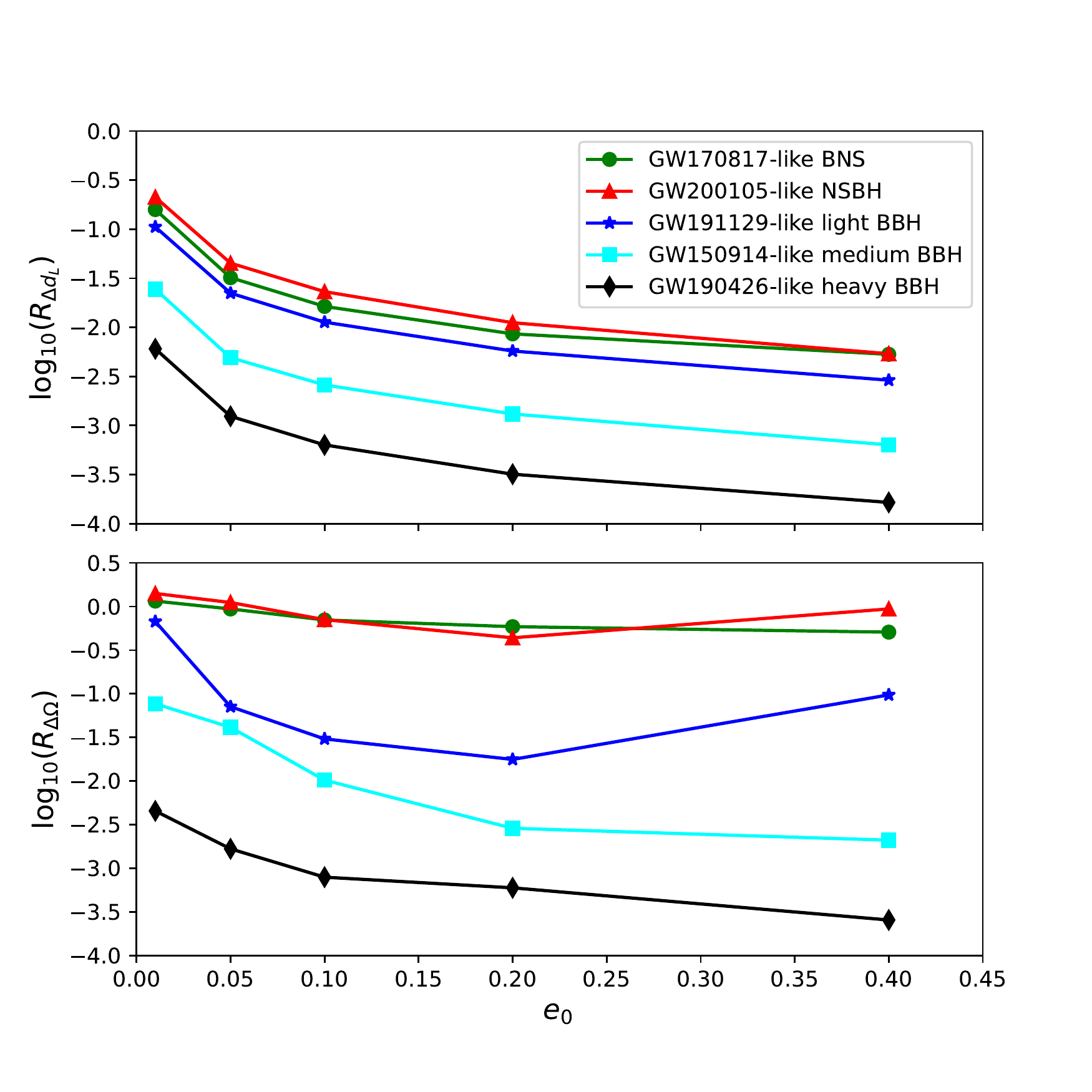} 
\caption{The largest improvement of distance inference (upper panel) and source localization (lower panel) for the typical compact binaries in 1000 orientations with variable eccentricities.}
\label{fig:Rwe}
\end{figure}

To illustrate the improvement of distance inference and localization for these typical binaries with variable eccentricities, we show the largest improvement ($\min(R)$ in 1000 orientations) of each case in Fig.~\ref{fig:Rwe}. For the distance inference, we can clearly see the trend of the improvement with the component mass and eccentricities of the binaries. Generally, a heavier binary with higher eccentricity can achieve more improvement of distance inference,  except for the BNS case which performs slightly better than the NSBH case. With eccentricity $e_0=0.4$, these typical binaries can achieve $\mathcal{O}(10^2-10^4)$ improvement (from lightest BNS to heaviest BBH) for the distance inference, compared to the circular case. For the source localization, the situation is more complicated. The effect of eccentricity is more distinct for a heavier binary. For BNS and NSBH cases, eccentricity does not improve the source localization obviously. Some eccentricities can even worsen the localization. While for the typical BBH cases, a nonvanishing eccentricity ($0.01\sim0.4$) can significantly improve the source localization, most by a factor of $1.5\sim{3.5}$ orders of magnitude. In addition, we note that for the light BBH case, enlarging eccentricity from 0.2 to 0.4 does not improve but worsens the localization. But for the medium and heavy BBH, they always benefit more from larger eccentricities. 
The reasons for the nontrivial features in the distance inference and source localizations are as follows.
On the one hand, eccentricity adds more harmonics to the waveform hence improving the SNR and breaking the parameter degeneracy, which can improve the parameter estimation. The higher harmonics which enter the detector band earlier could also provide more angular information. On the other hand, eccentricity reduces the observational time in the specific frequency band, which, however, lower the SNR and worsen the parameter inference and source localization. Because of the different starting frequency $f_{\rm start}$, the frequency ranges of the multiple harmonics of these typical binaries covered in the detector band (0.1-10 Hz) are also different. Moreover, two additional free parameters in eccentric waveform may also degrade the parameter inference. These factors compete with each other and make the parameter inference differ from case to case.

\textit{Conclusion and discussion.}--
To make dark sirens be the accurate and precise probes of the Universe, the distance inference and source localization are two crucial factors. GWs of the long inspiraling compact binaries observed by the space-borne detector in the mid-frequency band can provide much tighter constraints in these two aspects, compared to the LIGO-Virgo band. Without EM counterparts, dark sirens in the mid-band still face the issues like large degeneracy between luminosity distance and inclination angle in the near face-on orientations, as well as the uncertain localization for the larger-mass BBH. In this Letter, we demonstrate that the eccentricity, which is more likely to be nonvanishing in the mid-band than in the LIGO-Virgo band, can greatly alleviate these issues. The eccentricity itself can also be constrained very tightly ($\Delta e_0\sim\mathcal{O}(10^{-6})$ when $e_0=0.01$) in the mid-band. For low-mass binaries like BNS and NSBH, eccentricity does not help for the localization. But we find that the source can be localized very precisely even in the circular case, with $\mathbb{E}(\Delta \Omega)\sim\mathcal{O}(10^{-6}\sim10^{-5})$ deg$^{2}$. For larger binaries like the typical medium BBH, with eccentricity $e_0=0.1$ the localization can be improved from $10^{-1}$ to $10^{-3}$ deg$^2$. This means, in the mid-band, for either small or large-mass binaries with eccentricity, such precise localization makes it possible to identify the unique host galaxy of the dark sirens. Besides, the precise localization can provide an early warning for the follow-up observation of GWs in the high-frequency band,  as well for the search of EM counterparts. As shown in Fig~\ref{fig:result}, the localization is more precise in the near face-on orientations where the eccentricity happens to significantly improve the distance inference. In addition, one of the main targets for the mid-band detector is the intermediate mass black hole binaries (IMBHs), while in this Letter, we showed that among the typical events the GW190426-like heavy BBH can benefit most from the eccentricity. Our results indicate that the eccentricity of the long inspiraling compact binaries in the mid-band is a perfect ingredient for dark sirens as precise probes of the Universe.

To preliminarily assess the potential of eccentric dark sirens on cosmology, we also simulate the population of eccentric BNS, NSBH, and BBH based on current knowledge. We follow~\cite{Belgacem:2019tbw,Yang:2021qge,Yang:2021xox} and set the local rate of BNS, NSBH, and BBH to be consistent with GWTC-3~\cite{LIGOScientific:2021psn}. We find that the population of eccentric GWs detected by DECIGO accumulates quickly in the redshift range $z\sim[0,5]$, with distribution peaking around $z\sim1.5-2$ and total number $\sim\mathcal{O}(10^4-10^5)$ for 1-year observation. It can even observe sparse events at $z>10$. We also find that the distance of the five typical eccentric events can be measured at 0.1\%-10\% level. If we go to further distance, DECIGO can even measure the distance for medium and heavy BBH at 1\% precision level up to redshift $z\sim3$. 

Our results imply the great potential of eccentric dark sirens on cosmic expansion history (Hubble constant), dynamics of dark energy, and gravity theory (by comparing GW luminosity distance to EM luminosity distance~\cite{Belgacem:2017ihm,Belgacem:2019tbw,Finke:2021aom,Yang:2021qge}) up to very high redshift. 
Note our important findings in this Letter do not rely on a specific detector like DECIGO or BBO, they are the general physical implications of GWs emitted by the long inspiraling eccentric compact binaries. Our study has great significance for GWs on detection, data analysis, and physics including cosmology, astrophysics, and fundamental physics.


\begin{acknowledgments}
This work is supported by National Research Foundation of Korea 2021M3F7A1082056.
R.G.C is supported by the National Natural Science Foundation of China Grants No. 11821505, No. 11991052, No. 11947302, No. 12235019 and by the National Key Research and Development Program of China Grant No. 2020YFC2201502. 
Z.C is supported in part by the National Key Research and Development Program of China Grant
No. 2021YFC2203001, in part by the NSFC (No.~11920101003 and No.~12021003), in part by ``the Interdiscipline Research Funds of Beijing Normal University'' and in part by CAS Project for Young Scientists in Basic Research YSBR-006.
\end{acknowledgments}
\bibliography{ref}

\end{document}